\documentstyle[aps,12pt,draft]{revtex}

\begin{document}
\title{Time Interval Operators}
\author{M. Ruzzi}
\address{Departamento de F\'{\i}sica \\
Universidade Federal de Santa Catarina\\
88040-900 Florian\'{o}polis \\
Santa Catarina -- Brazil}
\author{D. Galetti}
\address{Instituto de F\'{\i}sica Te\'{o}rica\\
Universidade Estadual Paulista\\
Rua Pamplona 145\\
01405-900 S\~{a}o Paulo -- SP -- Brazil}
\maketitle

\begin{abstract}
In this work we will advance farther along a line previously developed
concerning our proposal of a time interval operator, on finite dimensional
spaces. The time interval operator is Hermitian, and its eigenvalues are
time values with a precise and interesting role on the dynamics. With the
help of the Discrete Phase Space Formalism (DPSF) previously developed, we
show that the time interval operator is the complementary pair of the
Hamiltonian. From that, a simple system is proposed as a quantum clock. The
only restriction is that our results do not apply to all possible
Hamiltonians. 
\end{abstract}


\section{Introduction}

In quantum mechanics the concept of time has been an issue always submitted
to continuous scrutiny and discussion, specially if one aims to treat it as
an observable. Although very interesting results have been already achieved,
the question is still far from to be closed. As it was shown long ago by
Pauli\cite{pauli}, direct attempts to find an operator which obeys a
canonical commutation relation with the Hamiltonian and, at the same time,
that have a reasonable physical interpretation are bounded to fail. In fact,
there are in the literature not too many proposals of Hermitian operators
which can be somehow related to time.

In this connection there are interesting approaches to the question of time
as an observable in quantum mechanics which should be mentioned. For
example, the method which takes benefit of a doubled Hilbert space \cite
{rosen,newt,bauer,lokun}, while others explore the possibility of defining
observables as positive-operator-valued (POV) measures\cite{busch}. In what
refers to the intriguing concept of time in the context of barrier
penetration there are several different approaches, although most of them do
not concern a time operator\cite{soko,pol,butt}, exception made to \cite
{kobe}.

Notwithstanding all this interesting approaches, there seems not to be an
attempt which can, at the same time, display a known algebra satisfied by
the Hamiltonian and the `time operator', with a physical interpretation for
all quantities involved, and with the time operator being an orthodox
Hermitian observable. From another point of view, we must also mention the
work by Pegg \cite{pegg}, which looks for a complementary operator to the
Hamiltonian, that have some results similar to those of this paper. In the
following we shall develop an idea which we recently proposed\cite{garu},
that fulfills all the conditions mentioned above provided that the physical
systems under consideration satisfy given conditions.

The starting point of our approach is based on the consideration of quantum
degrees of freedom characterized by finite dimensional state spaces. As has
been shown, in these cases it is possible to make use of the Discrete Phase
Space Formalism (DPSF) as a suitable auxiliary tool\cite{gapi,gapi2,ruga}
since this formalism can be seen as a generalization of the Weyl-Wigner
formalism to degrees of freedom without classical counterpart. As such,
operators acting on the appropriate Hilbert (ket) spaces can be mapped onto
functions of integer indices provided that a basis in the operator space is
well defined. By its turn, it has been proposed that the functional form of
the operator basis elements consist, in general terms, of a double discrete
Fourier transform of the Schwinger operator basis\cite{Schw}. Once the
action of the operator basis elements on relevant state kets is defined, the
mapping procedure, that takes operators to the corresponding representative
function in the discrete phase space, can be put to practical terms. In
section II we briefly review the main ideas referring to the operator basis.

Since we dispose of a mapping procedure obtained from the DPSF, operator
equations can be converted into ordinary equations involving the phase space
counterparts (the mapped expressions) of the respective operators. If we
restrict ourselves to degrees of freedom of the type we are considering,
with no classical counterpart, this convenient feature permits us to show
that it is possible to find an operator, which we call time interval
operator, that is a generator of energy shifts and its exponential obeys a
Weyl-Schwinger commutation relation with the time evolution operator. This
particular operator, defined for such systems, is a Hermitian operator such
that its spectrum is a discrete set of time intervals with an important role
in the dynamical evolution. This is, however, only accomplished by a certain
class of Hamiltonians, and the Weyl-Schwinger algebra is fulfilled only for
some prescribed time intervals, which are in fact the eigenvalues of the
time interval operator. In this connection we shall see that there is no
general time interval operator but rather a different time interval operator
for each particular Hamiltonian which admits such a `canonical' pair. The
time evolution governed by this particular time interval operator,
satisfying that algebra, characterizes a particular dynamics since it
evolves the state of the system over the sites of the discrete phase space
occupying only one site (exhausting the total probability) at each time
interval, and must not be confused with the general time evolution governed
by the usual continuous time, which, by its turn, can always be carried out.
This discussion is basically what is accomplished in sections III and IV.

Although the proposal of a discrete time has been discussed in other
contexts, in this work we are not making any hypothesis about the background
c-number valued parameter $t$, but rather discussing the properties of an
operator and its eigenstates, a set of discrete time intervals which we
denote `quantum clock times', as we propose a simple quantum clock in the
end of section IV. The move from the independent variable time to the
concept of time interval is the key to the whole process, that in Section V
is presented with some general remarks and conclusions.

\section{The Operator Basis}

Considering a time independent Hamiltonian and its set of eigenstates
consisting of an $N-$dimensional ket space, 
\begin{equation}
H|u_k\rangle =E_k|u_k\rangle ,\qquad k=0,1,...N-1,  \label{2}
\end{equation}
we construct{\em \ }the operator 
\begin{equation}
V=\sum_{k=0}^{N-1}|u_k\rangle \langle u_{k+1}|,  \label{3}
\end{equation}
where hereafter we adopt a cyclic notation 
\begin{equation}
|u_k\rangle \equiv |u_{k(%
\mathop{\rm mod}
N)}\rangle  \label{3.3}
\end{equation}
which in particular means that in Eq.(\ref{3}) the last term of the
summation is in fact $|u_{N-1}\rangle \langle u_0|.$ From this definition it
follows that 
\begin{equation}
V^s|u_n\rangle =|u_{n-s}\rangle .  \label{3.6}
\end{equation}
Now the operator $V$ can in turn be diagonalized,and once its eigenstatesare
obtained, it can be directly verified that the well known Schwinger unitary
operators algebra results are reproduced \cite{Schw}, that is,

\begin{equation}
U|u_{k}\rangle =\exp \left[ \frac{2\pi i}{N}k\right] |u_{k}\rangle ,\qquad
V|v_{k}\rangle =\exp \left[ \frac{2\pi i}{N}k\right] |v_{k}\rangle ,
\label{4}
\end{equation}
\begin{equation}
U^{N}=V^{N}=\hat{1},  \label{unt}
\end{equation}
\begin{equation}
V^{s}|u_{n}\rangle =|u_{n-s}\rangle ,\qquad U^{s}|v_{n}\rangle
=|v_{n+s}\rangle ,  \label{5}
\end{equation}
being that in our case the set $\left\{ |u_{n}\rangle \right\} $
diagonalizes both the Hamiltonian and the Schwinger operator $U$.
Furthermore, the two sets of states are connected by a discrete Fourier
transform 
\begin{equation}
\langle v_{k}|u_{n}\rangle =\frac{1}{\sqrt{N}}\exp \left[ -\frac{2\pi i}{N}%
kn\right] ,  \label{8}
\end{equation}
and the operators obey the Weyl-Schwinger algebra 
\begin{equation}
U^{j}V^{l}=\exp \left[ \frac{2\pi i}{N}jl\right] V^{l}U^{j}.  \label{9}
\end{equation}

It has been also shown that the set of $N^2$ operators\cite{ruga} 
\begin{equation}
\hat{G}(m,n)=\frac 1N\sum_{j,l=-\frac{N-1}2}^{\frac{N-1}2}U^jV^l\exp \left[ 
\frac{\pi i}Njl\right] \exp \left[ -\frac{2\pi i}N(mj+nl)\right]  \label{10}
\end{equation}
constitutes a complete and orthogonal basis in operator space; the relevant
properties of this basis are collected in reference \cite{ruga}, from which
we recall the main results. In the form that it is presented, this basis is
suited to deal with odd dimensional spaces, although it can be easily
adapted to even ones. In fact, as this basis keeps, from the Schwinger
basis, the property of decomposition into independent sub-bases, each with a
prime dimension\cite{Schw}, we, for simplicity, shall hereafter consider
only prime $N$'s, although most of the results in what follows do not depend
on this condition. We are also omitting the so called modular phase, which
should, in principle at least, be present in the basis, but is irrelevant in
the particular present calculations.

Thus, any operator acting on those finite dimensional state spaces can be
written in the form 
\begin{equation}
\hat{O}=\frac{1}{N}\sum_{j,l=-\frac{N-1}{2}}^{\frac{N-1}{2}}o(m,n)\hat{G}%
(m,n),  \label{10.3}
\end{equation}
where the coefficients (the operator representative on the discrete phase
space, or equivalently, the discrete Weyl-Wigner mapped function associated
to the operator) are given by 
\begin{equation}
o(m,n)=Tr[\hat{G}^{\dagger }(m,n)\hat{O}].  \label{10.6}
\end{equation}

\section{Time evolution as a lattice shift}

Systems with a time independent Hamiltonian $H$ have a time evolution
operator of the form 
\begin{equation}
{\cal G}(\Delta t)=\exp \left[ \frac{-iH\Delta t}\hbar \right] ,  \label{11}
\end{equation}
that propagates the states $|\psi (t)\rangle $, to which corresponds the
density operator $\hat{P}(t)=|\psi (t)\rangle \langle \psi (t)|$ for a
general pure state (although this is not an essential assumption for what
follows) in such a way that 
\begin{equation}
\hat{P}(t+\Delta t)={\cal G}(\Delta t)\hat{P}(t){\cal G}^{\dagger }(\Delta
t).  \label{12}
\end{equation}

Now let us map this operator equation onto the discrete phase space assuming
the {\em hypothesis} 
\begin{equation}
{\cal G}(\Delta \tau )=U^{-k},\;\;\;\;k\in [0,N-1],  \label{hyp}
\end{equation}
that is, the unitary time evolution operator is equal to some integer power
of the Schwinger unitary operator $U$, at least for some given time interval 
$\Delta \tau $. A negative power is chosen as a matter of convenience,
recalling that, according to Eq. (\ref{unt}), $U^{-k}=U^{N-k}$. Fairness of
such hypothesis shall be justified {\it a posteriori}.

Hypothesis (\ref{hyp}) has a direct consequence. If it holds for a given $%
\Delta \tau $, then it follows that, for any integer $n$, 
\begin{equation}
\left( \exp \left[ -\frac{iH\Delta \tau }\hbar \right] \right) ^n=\left(
U^{-k}\right) ^n  \label{19}
\end{equation}
\begin{equation}
\exp \left[ -\frac{iHn\Delta \tau }\hbar \right] =U^{-kn},  \label{20b}
\end{equation}
and cyclicity of $U$ implies that 
\begin{equation}
\exp \left[ -\frac{iHn\Delta \tau }\hbar \right] =U^{-kn\;\;(\mathop{\rm mod}%
N)},  \label{20}
\end{equation}
which obviously means that if Eq. (\ref{hyp}) holds for a given time
interval, a similar equation must hold for all its integer multiples. In
view of that we can rewrite our hypothesis as 
\begin{equation}
{\cal G}(\Delta \tau _n)=U^{-k^{\prime }},\;\;\;\;k^{\prime }\in [0,N-1],
\label{hyp2}
\end{equation}
where 
\begin{equation}
\Delta \tau _n=n\Delta \tau ,\;\ \;\;n=1,2,3...,
\end{equation}
and 
\begin{equation}
k^{\prime }=kn\;\;(\mathop{\rm mod}N),
\end{equation}
what means that hypothesis (\ref{hyp}) refers in fact to an infinite
sequence of time intervals and not just to a given one.

A less straightforward consequence can be extracted with the help of the
discrete phase space representation of the density operator which is, as in
the continuous cases, called Wigner function and denoted by ${\rho }%
_w(m,n;t) $. This function can be immediately calculated from 
\begin{equation}
{\rho }_w(m,n;t+\Delta \tau )=Tr[G^{\dagger }(m,n)\hat{P}(t+\Delta \tau )],
\label{33b}
\end{equation}
and if one assumes that the hypothesis (\ref{hyp}) is true, then (using Eq. (%
\ref{12})) 
\begin{equation}
{\rho }_w(m,n;t+\Delta \tau )=Tr[G^{\dagger }(m,n)U^{-k}\hat{P}(t)U^k].
\label{33}
\end{equation}
Now, if we substitute the density operator at time $t$ by the expression
giving its decomposition in the operator basis, 
\begin{equation}
{\rho }_w(m,n;t+\Delta \tau )=Tr[\hat{G}^{\dagger }(m,n)U^{-k}\left( \frac 1N%
\sum_{r,s=-h}^h{\rho }_w(r,s;t)\hat{G}(r,s)\right) U^k]  \label{35}
\end{equation}
\[
{\rho }_w(m,n;t+\Delta \tau )=Tr[\sum_{j,l=-h}^h\frac 1NV^{-l}U^{-j}\exp
\left[ -\frac{i\pi }Njl\right] \exp \left[ \frac{2\pi i}N(mj+nl)\right]
\times 
\]
\begin{equation}
U^{-k}\left( \frac 1N\sum_{r,s=-h}^h{\rho }_w(r,s;t)\sum_{x,z=-h}^h\frac 1N%
U^zV^x\exp \left[ \frac{i\pi }Nxz\right] \exp \left[ -\frac{2\pi i}N%
(xr+zs)\right] \right) U^k].  \label{36}
\end{equation}
Using the Weyl commutation relation between $V^x$ and $U^k$, 
\[
{\rho }_w(m,n;t+\Delta \tau )=\frac 1{N^3}\sum_{j,l=-h}^h\sum_{r,s=-h}^h%
\sum_{x,z=-h}^hTr[V^{-l}U^{-j}U^zV^x]\exp \left[ -\frac{i\pi }Njl\right]
\exp \left[ \frac{2\pi i}N(mj+nl)\right] 
\]
\begin{equation}
\exp \left[ -\frac{2\pi i}Nxk\right] {\rho }_w(r,s;t)\exp \left[ \frac{i\pi }%
Nxz\right] \exp \left[ -\frac{2\pi i}N(xr+zs)\right] .  \label{37}
\end{equation}
The trace contribution on the r.h.s. of (\ref{37}) is $N\delta
_{z,j}^{[N]}\delta _{x,l}^{[N]}$, where $\delta _{a,b}^{[N]}$ is a Kronecker
delta modulo $N$ ({\it i.e.}, it is only different from zero if $n=i$ $(%
\mathop{\rm mod}N)$), and then

\begin{equation}
{\rho }_{w}(m,n;t+\Delta \tau )=\frac{1}{N^{2}}\sum_{j,l=-h}^{h}%
\sum_{r,s=-h}^{h}\exp \left[ \frac{2\pi i}{N}(j(m-r)+l(n-s-k))\right] {\rho }%
_{w}(r,s;t).  \label{39}
\end{equation}
We first observe that the sums over $\{j,l\}$ also yield Kronecker deltas $%
\mathop{\rm mod}
N$, in fact $N\delta _{m,r}^{[N]}\delta _{n-k,s}^{[N]}$, and after the sums
over $\{r,s\}$ are performed we finally obtain 
\begin{equation}
{\rho }_{w}(m,n;t+\Delta \tau )={\rho }_{w}(m,n-k;t).  \label{41}
\end{equation}

Equation (\ref{41}) shows that hypothesis (\ref{hyp}), for that precise time
interval for which it holds, leads the system state to a situation
equivalent to a shift of $k$ sites along one direction in the
two-dimensional discrete phase space. Use of equation (\ref{hyp2})
characterizes a time evolution, since for each time interval there will be
associated a given shift in phase space. A particularly simple and
elucidative physical situation where this occurs is presented in reference 
\cite{garu}.

We are yet to determine the physical conditions which allow our hypothesis
to be valid. In order to achieve that we make use of the discrete phase
space representation of the operator equation proposed as our hypothesis.
The mapped expressions of the individual operators read 
\begin{equation}
(U^{-k})(m,n)=\exp \left[ -\frac{2\pi i}Nkm\right] ,  \label{17}
\end{equation}
\begin{equation}
({\cal G}(\Delta t))(m,n)=g(m,n;\Delta t)=\exp \left[ -\frac{iE_m\Delta t}%
\hbar \right] ,  \label{18}
\end{equation}
respectively, so that on the discrete phase space Eq. (\ref{hyp}) will be
written as

\begin{equation}
\exp \left[ -\frac{iE_m\Delta t}\hbar \right] =\exp \left[ -\frac{2\pi i}N%
km\right] ,  \label{24}
\end{equation}
and therefore 
\begin{equation}
\frac{E_m\Delta t}\hbar =\frac{2\pi }Nkm\quad (%
\mathop{\rm mod}
2\pi ).  \label{25}
\end{equation}
Equation (\ref{25}) is in fact a set of $N$ equations (one for each $m$)
which must hold separately for a given $\Delta t$ {\em and} for a given $k$.
It can be still written as 
\begin{equation}
E_m\frac{N\Delta t}{2\pi \hbar }=km+Nf(m).  \label{25.7}
\end{equation}
where $f(m)$ is an arbitrary integer function of an integer variable. The
r.h.s. of Eq. (\ref{25.7}) is an integer number, whereas the l.h.s is a
product of two, in principle, real numbers. It is then clear that a
necessary but not sufficient condition to Eq. (\ref{25.7}) holds is that
there must be a real number $\lambda $ for which the whole set $\left\{ 
\frac{E_m}\lambda \right\} $ is an integer. If we want that condition to
become sufficient, the set must be additionally a complete set of remainders
modulo $N$\cite{andrews}. Summarizing, Eq. (\ref{24}) will only have
solutions if and only if the spectrum can be written as 
\begin{equation}
E_m=\hbar \omega \left( km+\ Nf(m)\right) ,  \label{26}
\end{equation}
where $\omega $ is an appropriately defined constant. From (\ref{25}) and (%
\ref{26}) it follows that the smallest time interval for which ${\cal G}%
(\Delta t)=U^{-k}$ is 
\begin{equation}
\Delta \tau =\frac{2\pi }{N\omega }  \label{27}
\end{equation}
and for all integer multiples of $\Delta \tau $ the time evolution operator
will be equal to a given power of $U.$ Once $N$ is a prime number, the set
of the first $N$ powers of $U^{-k}$ will be equivalent to the set $%
\{1,U,U^2,..,U^{N-1}\}$ in some permuted order. Therefore, all powers of $U$
(and consequently the respective sites of the discrete phase space) are
visited as the time intervals are counted.

\section{Time Interval Operator}

We have noticed that, for Hamiltonians which fulfill Eq. (\ref{26}) and on
time intervals which are multiples of (\ref{27}), the time evolution
operator is equal to an integer power of the Schwinger unitary operator $U$.
If that is true, the Weyl commutation relation can be written as 
\begin{equation}
\exp \left[ \frac{-iH(n\Delta \tau )}\hbar \right] V^{-j}=\exp \left[ \frac{%
2\pi i}Nj(nk)\right] V^{-j}\exp \left[ \frac{-iH(n\Delta \tau )}\hbar
\right] .  \label{43}
\end{equation}
On the other hand, we can write the Schwinger operator $V$ itself as the
exponential of an operator if we define 
\begin{equation}
{\rm T}=\frac{2\pi }{\omega N}\sum_{j=0}^{N-1}j|v_j\rangle \langle v_j|,
\label{44}
\end{equation}
as it can be directly verified that 
\begin{equation}
\exp \left[ -\frac{i{\rm T}(E_{m+s}-E_m)}\hbar \right] =V^{-ks},  \label{45}
\end{equation}
once the energy spectrum satisfies (\ref{26}). Equation (\ref{43}) therefore
can be written as 
\begin{equation}
\exp \left[ \frac{-iHn\Delta \tau }\hbar \right] \exp \left[ \frac{-i{\rm T}%
\Delta E_j}\hbar \right] =\exp \left[ \frac{2\pi i}Nnkjk\right] \exp \left[ 
\frac{-i{\rm T}\Delta Ej}\hbar \right] \exp \left[ \frac{-iHn\Delta \tau }%
\hbar \right] ,  \label{46}
\end{equation}
where $\Delta E_j=E_{s+j}-E_s.$

From Eq. (\ref{44}) it is not difficult to obtain the eigenstates and
eigenvalues of ${\rm T}$, namely

\begin{equation}
{\rm T}|v_{l}\rangle =t_{l}|v_{l}\rangle =\Delta \tau l|v_{l}\rangle ,
\label{47}
\end{equation}
and its phase space representative 
\begin{equation}
\left( {\rm T}\right) (m,n)=\Delta \tau n.  \label{48}
\end{equation}

\subsection{Quantum Clock}

Let a given physical system $S$, described by a Hamiltonian whose spectrum
obeys Eq. (\ref{26}), be, at an instant $t_0,$ in an eigenstate $|v_i\rangle 
$ of the time interval operator. Its corresponding initial discrete Wigner
function is 
\begin{equation}
\rho _w(m,n;t_0)=Tr[G^{\dagger }(m,n)|v_i\rangle \langle v_i|]  \label{q56}
\end{equation}
which can be easily calculated as 
\begin{equation}
\rho _w(m,n;t_0)=\delta _{n,i}^{[N]}.  \label{q57}
\end{equation}
Using the result of Eq. (\ref{41}) for the time evolution 
\[
{\rho }_w(m,n;t_0+\Delta \tau )={\rho }_w(m,n+k;t_0) 
\]
it is simple to verify that 
\[
{\rho }_w(m,n;t_0+\Delta \tau )=\delta _{n,i+k}^{[N]}, 
\]
which can be generalized, through the use of (\ref{hyp2}) to 
\[
{\rho }_w(m,n;t_0+j\Delta \tau )=\delta _{n,i-jk}^{[N]}. 
\]
This is a quite interesting result. It simply states that, when the system
starts its evolution at (or visit in a given instant) an eigenstate of the
time interval operator, it will be in another eigenstate, with {\em no
uncertainty,} after every time interval $\Delta \tau .$ All eigenstates
shall be visited in the time interval $(N-1)\Delta \tau $, although, if $%
k\ne 1$, they may not appear in the correct order. In general the sequence
will be 
\begin{equation}
|v_i\rangle ,|v_{i+k}\rangle ,|v_{i+2k}\rangle ......|v_{i+(N-1)k}\rangle ,
\end{equation}
where the reader should remind of the modulo $N$ extraction in the subscript
of the ket labels. In view of that, the next ket in the above list would be 
\begin{equation}
|v_{i+Nk}\rangle \equiv |v_i\rangle ,
\end{equation}
and the system would be back to where it started, that is, within such a
cycle, the index of state ket is a 'counter' of time.

This strongly motivates us to denote this dynamically determined $\Delta
\tau $ (and its first $(N-1)$ integer multiples) as quantum clock time.

\section{Conclusions}

The basic result that we have achieved here is a direct consequence of the
quantum mechanics associated to degrees of freedom characterized by a finite
dimensional state space and from trying to give quantum mechanical meaning
to time intervals rather than to time itself. We have identified a Hermitian
operator, a quantum mechanical observable, which is directly related to time
intervals. We have denoted this operator as time interval operator, and we
stress again that it is not directly identified with the continuous usual
time parameter $t$. When equations (\ref{hyp}) and (\ref{26}) are
simultaneously satisfied we are dealing with physical systems such that this
operator displays the following properties:

\begin{itemize}
\item[i)]  it is a generator of cyclical shifts in energy state space;

\item[ii)]  its exponential obeys the Weyl-Schwinger commutation relation
together with the time evolution operator;

\item[iii)]  it is a Hermitian operator whose eigenstates form a set of time
intervals with an important role in dynamics;

\item[iv)]  the eigenstates of the time interval operator are connected with
the energy eigenstates through a discrete Fourier transform, therefore these
two sets have a maximum degree of incompatibility.
\end{itemize}

The Weyl-Schwinger commutation relation, however, is only fulfilled for some
given values of time intervals, exactly the eigenvalues of the time interval
operator. The time evolution for which the Weyl-Schwinger commutation
relations hold could then be called stroboscopic, in the sense that, in this
case, during this time evolution, it always happens that each one of the
sites of the energy sector of the discrete phase space individually will be
visited and will exhaust the total probability when the time evolves as
multiples of a given time interval (which are the eigenvalues of the time
interval operator). By its turn, this particular dynamics must not be
confused with that governed by the usual continuous time parameter, which
can always be also carried out. Therefore, following the algebraic structure
of those operators and their corresponding spectra, the equations
characterizing the time interval operator allowed us to classify some simple
physical systems that behave as a quantum clock.

In this connection, let us discuss some basic operational requirements
related to the interpretation of the concept of time. To measure time, there
must be a reference to a nonstationary quantity -- or a property -- of a
physical system, and, in general, this abstract quantity admittedly changes
continuously with time. From an orthodox quantum mechanical point of view, a
measurement is related to a Hermitian operator, and the process of measuring
may destroy the clock itself. This consideration illustrates the importance
of the identification of a Hermitian operator related to time, and why we
have called our time intervals of `clock times', since a measure would only,
in the worst case, 'reset' the clock. We, in fact, have not identified a
continuous changing property but rather a discrete changing property of a
physical system to serve as a `quantum clock'. This comes as a natural
feature of the finite dimensional spaces we are dealing with.

\end{document}